# Updated constraint for the gravitational wave background from the gamma-ray Pulsar Timing Array

Serena Valtolina,[1,2,3,*] Colin J. Clark,[1,2] Rutger van Haasteren,[1,2] Aurélien Chalumeau,[4,5] H.Thankful Cromartie,[6,†] Matthew Kerr,[7] Lars Nieder,[1,2] and Aditya Parthasarathy[4,8,3]

[1]*Max Planck Institute for Gravitational Physics (Albert Einstein Institute), D-30167 Hannover, Germany*
[2]*Leibniz Universität Hannover, D-30167 Hannover, Germany*
[3]*Max-Planck-Institut für Radioastronomie, Auf dem Hügel 69, 53121 Bonn, Germany*
[4]*ASTRON, Netherlands Institute for Radio Astronomy, Oude Hoogeveensedijk 4, 7991 PD Dwingeloo, The Netherlands*
[5]*LPC2E, OSUC, Univ Orleans, CNRS, CNES, Observatoire de Paris, F-45071 Orleans, France*
[6]*National Research Council Research Associate, National Academy of Sciences, Washington, DC 20001, USA*
[7]*Space Science Division, Naval Research Laboratory, Washington, DC 20375–5352, USA*
[8]*Anton Pannekoek Institute for Astronomy, University of Amsterdam, Science Park 904, 1098 XH Amsterdam, The Netherlands*

Fermi Large Area Telescope observations of gamma-ray pulsars can be used to build a pulsar timing array (PTA) experiment to search for gravitational wave (GW) signals at nanohertz frequencies. At those frequencies, the dominant signal is expected to be a stochastic gravitational wave background (GWB) produced by the incoherent superposition of the quasimonochromatic GW emissions from a population of supermassive black hole binaries. While the radio PTAs have recently announced compelling evidence for a GWB signal with a power law spectrum of strain amplitude $\approx 2\text{–}3 \times 10^{-15}$ (at the frequency of 1 $\mathrm{yr}^{-1}$), in 2022, an analysis of 12.5 years of Fermi data for 35 pulsars led to an upper limit of $1 \times 10^{-14}$ for the GWB amplitude. The analysis was carried out on times of arrival obtained by folding from six months up to one year of photon observations. A *photon-by-photon* approach was also tested to infer constraints on the GWB amplitude from individual pulsars, but without accounting for the cross-pulsar correlations that a GWB would induce. Here, we reanalyze the same dataset using a regularized likelihood method that correctly models cross-pulsar correlations directly from the photons, while additionally marginalizing over the uncertain pulse profile shape. While the two methods are not expected to have significant differences in sensitivity, we prove through simulations of gamma-ray PTA datasets that the *photon-by-photon* method for GWB recoveries is, statistically, more robust. The resulting upper limit obtained for the GWB strain amplitude is $1.2 \times 10^{-14}$, indicating that the improved method yields a consistent result with the previous analyses.

## I. INTRODUCTION

Timing a pulsar means constructing a mathematical *timing model* which accurately predicts the observed times of arrival (TOAs) of the pulsar's emission as a function of some parameters $\boldsymbol{\beta}$ describing the pulsar's rotation, signal propagation and the relative motion of the pulsar and Earth [1]. The TOAs recorded for an *array* of pulsars are the primary data of pulsar timing array (PTA) experiments.

Timing residuals are the extra delays which are recorded in pulsar observations compared to those predicted by the timing model. These delays are tentatively explained by a combination of different factors, such as intrinsic instabilities in the pulsar's emission mechanism, interaction of the propagating signal with the interstellar medium, dynamics of the pulsar's magnetosphere, and more. While these signals carry information about the properties of individual neutron stars, they also collectively encode a delay induced by a nanohertz gravitational wave (GW) signal, expected to be dominated by a stochastic GW background (GWB), due to the incoherent superposition of quasimonochromatic GW emission from a population of supermassive black hole binaries (SMBHBs) [2–6]. A GWB is usually modeled as a common low-frequency signal that induces correlated delays in the pulsar TOAs. This correlation between the timing residuals of different pulsars depends only upon the pulsars' angular separation and is predicted by the Hellings and Downs (HD) correlation function [7,8]. Characterizing the GW-induced delays in pulsar observations is the main goal of PTA experiments [9].

[*]Contact author: svaltolina@mpifr-bonn.mpg.de
[†]Present address: Resident at Naval Research Laboratory, Washington, DC 20375, USA.

Pulsar timing is primarily done using radio telescopes with large collecting areas, which can measure the pulse arrival times from millisecond pulsars with precision of $\approx 1\ \mu s$ in observation times of $\approx 1$ ks. In the radio band, continuous observations are folded to produce high signal-to-noise (S/N) pulse profiles by averaging over many spin periods. This nominal pulse profile is then compared to a high S/N template pulse profile to determine a precise TOA according to the telescope's atomic clock [1]. Despite the great precision of the measured radio TOAs, pulse propagation is strongly affected by interaction with the interstellar medium (the so-called dispersion measure, DM, effects), solar wind effects, pulse variability, scattering and more (see, for example, Refs. [10–15]). Each of these phenomena appears in the data as a stochastic signal which often has amplitude comparable to, if not higher than, the sought GWB signal. This makes the radio pulsar noise models highly complicated and characterized by many covariant signals.

It is also possible to time pulsars at higher energies, including with gamma rays [16]. In particular, the Fermi Large Area Telescope (Fermi LAT) [17] records individual photons. While each is time stamped with $\approx 300$ ns precision [17], the low photon flux and typically broad pulse profiles mean that many months of data must be folded to obtain a pulsar TOA estimate with similar $\mu s$ precision.

The alternative is to consider each photon individually, without any folding procedure: a *photon-by-photon* approach, e.g., Ref. [18] that optimizes the Poisson likelihood of the entire set of photons, given a timing model and an assumed pulse shape. Offsetting the limited sensitivity, the noise model associated to the gamma-ray data is much simpler than the model required for radio data analysis. Because of the high frequency of the gamma rays, there is no interaction with the interstellar medium. Thus, noise models for gamma-ray pulsars include only the pulsar's spin noise and Poisson statistics noise. Furthermore, the gamma-ray pulsar data have an essentially constant experimental setup (Fermi LAT data are uninterrupted and calibrations have been constant for the full dataset) and they are potentially less subject to astrophysical effects such as changes in the pulse shape.

Recently, many PTA collaborations (the Chinese PTA [CPTA], the European PTA [EPTA] together with the Indian PTA [InPTA], NANOGrav, the Parkes PTA [PPTA] and the MeerKAT PTA collaborations) have reported evidence for a common red process[1] with cross-pulsar correlation properties consistent with the sought-after HD correlation [19–23]. The only gamma-ray PTA (GPTA) study published so far [24] sets an upper limit for the strain amplitude of a possible GWB signal to $1 \times 10^{-14}$ (normalized at the frequency of 1 $\mathrm{yr}^{-1}$). In that work, using a cadence up to one year, it was possible to compute well-defined TOAs only for 29 out of the 35 pulsars. No method was available for searches of correlated signals using the photon-by-photon approach from different pulsars.

In this paper, we present the first search for an HD-correlated GW signal directly from photon data (and the resulting upper limit on the GWB amplitude) for the 35 pulsars in Ref. [24] using the method introduced in Ref. [25]. This approach is based on a regularized formulation of the PTA likelihood in Fourier space that requires as input only the posterior distribution of the Fourier coefficients used to model the Gaussian processes involved (like intrinsic timing noise (TN) or the GWB). Here, we test the performance of the method on simulated GPTA data and show the results obtained for the GWB search from the photon data of Ref. [24].

The paper is structured as follows: in Sec. II, we present an overview of both the TOAs approach and the derivation of the regularized formulation of the PTA likelihood to carry out the GWB inference directly from the photon observations. We also present a summary of the results of the previous analyses on the first GPTA dataset in Sec. II C. Then, we test the sensitivity and accuracy of the method on simulated datasets in Sec. III. Finally, we present the results of our analysis of the 35 gamma-ray pulsars of Ref. [24] in Sec. IV and discuss future directions in Sec. V.

## II. METHODS

In this section, after a brief review of the Bayesian description used for timing residuals in radio PTA experiments and a summary of the results and methods applied in the first data release of the GPTA [24], we describe the formalism introduced in Ref. [25] for inference of a GW-correlated signal from pulsar gamma-ray photon observations. See Ref. [25] for a more detailed derivation of the method and its validation on radio PTA data.

### A. TOAs approach

Timing residuals $\delta \mathbf{t}$ are obtained from observed TOAs ($\mathbf{t}^{\mathrm{obs}}$) as $\delta \mathbf{t} \equiv \mathbf{t}^{\mathrm{obs}} - f(\mathbf{t}; \boldsymbol{\beta}_0)$, where the term $f(\mathbf{t}; \boldsymbol{\beta}_0)$ corresponds to the TOAs predicted by the timing model $f$ evaluated at the reference values $\boldsymbol{\beta}_0$ (usually best guesses from an initial timing analysis) for the timing parameters $\boldsymbol{\beta}$. $\delta\mathbf{t}$ can be rewritten as a sum of stochastic delay components.

[1]The adjective *red* means that the power spectral density is higher at lower frequencies. See Eq. (15).

We assume our timing model solution is accurate enough that extra delays due to timing model ephemeris offsets are well approximated by the first-order correction term of the linear expansion of $f(\mathbf{t};\boldsymbol{\beta})$ around the solution $\boldsymbol{\beta}_0$. Thus, we compute these additional delays as the product between the *design matrix* $M$, where $M_{ij} \equiv (\partial f(\mathbf{t}_i;\boldsymbol{\beta})/\partial\boldsymbol{\beta}_j)|_{\boldsymbol{\beta}_0}$, and the offsets $\boldsymbol{\xi} \equiv \boldsymbol{\beta} - \boldsymbol{\beta}_0$.

Other delays induced by low-frequency (red) processes, such as pulsar intrinsic spin noises, DM variations (due to interactions with the turbulent ionized interstellar medium), GWs and more, are modeled with a low-rank approximation to a Gaussian process that uses a Fourier basis over $k$ frequency components ($f_k \equiv k/T_{\mathrm{tot}}$ where $T_{\mathrm{tot}}$ is the total time of observation), as follows:

$$\sum_k [a_k \cos(2\pi k\mathbf{t}/T_{\mathrm{tot}}) + b_k \sin(2\pi k\mathbf{t}/T_{\mathrm{tot}})]\nu_{\mathrm{obs}}^{\alpha} = F\mathbf{a}, \quad (1)$$

where $F$ is called the *Fourier design matrix*, $\mathbf{a}$ is the array of all Fourier coefficients $(a_k, b_k)$, and the term $\nu_{\mathrm{obs}}^{\alpha}$ includes the dependence upon the observing frequency (for achromatic signals $\alpha = 0$). We also impose a prior $\phi$ for the distribution of $\mathbf{a}$ which is a function of the hyperparameters $\boldsymbol{\rho}$ that describe the power spectral density of the involved Gaussian processes.

For the model-subtracted residuals $\delta\mathbf{t} - T\mathbf{b}$ (where $T\mathbf{b} \equiv M\boldsymbol{\xi} + F\mathbf{a}$ for notation compactness), the likelihood function will be a Gaussian[2] with covariance $N$, which is a function of some noise parameters $\boldsymbol{\eta}$: $\mathcal{N}(\delta\mathbf{t}|T\mathbf{b}, N)$.

The full posterior distribution for the noise parameters $\boldsymbol{\eta}$ and hyperparameters $\boldsymbol{\rho}$ of an array of $N_p$ pulsars is written as

$$p(\mathbf{b},\boldsymbol{\eta},\boldsymbol{\rho}|\delta\mathbf{t}) = \left[\prod_{k=1}^{N_p} \frac{p(\delta\mathbf{t}_k|\mathbf{b}_k,\boldsymbol{\eta})}{p(\delta\mathbf{t}_k)}\right] p(\mathbf{b}|\boldsymbol{\rho})p(\boldsymbol{\rho})p(\boldsymbol{\eta}), \quad (2)$$

where $p(\delta\mathbf{t}_k)$ is the fully marginalized likelihood (often called evidence) for the timing residuals of the $k$th pulsar, $p(\boldsymbol{\eta})$ and $p(\boldsymbol{\rho})$ are the prior distributions for the noise parameters and hyperparameters, and $p(\mathbf{b}|\boldsymbol{\rho}) \equiv \mathcal{N}(\mathbf{b}|0, B)$. The matrix $B$ contains the prior for both the timing model ephemerides offsets and the Fourier coefficients. It is customary in PTA analysis to set an improper infinite prior on the timing model ephemeris offsets, since the data are highly informative for these parameters. Thus, $B$ is defined as a block diagonal matrix $B \equiv blockdiag(\infty, \phi)$.

Usually in PTA analysis for Bayesian inference on the GW hyperparameters [which enter Eq. (2) in the $B$ matrix: $B = B(\boldsymbol{\rho})$], we use the marginalized version of Eq. (2) over the timing model ephemerides and the Fourier coefficients, as follows:

$$p(\boldsymbol{\eta},\boldsymbol{\rho}|\delta\mathbf{t}) = \int \mathrm{d}\mathbf{b}\, p(\mathbf{b},\boldsymbol{\eta},\boldsymbol{\rho}|\delta\mathbf{t}) \propto \mathcal{N}(\delta\mathbf{t}|0, C)p(\boldsymbol{\eta})p(\boldsymbol{\rho}), \quad (3)$$

where $C = N + TBT^T$.

See Refs. [26–30] for more details on the PTA likelihood definition for the TOAs analysis.

### B. Regularized Fourier likelihood and GW search from photons

The work presented in Ref. [25] introduces a regularized formulation of the PTA likelihood that allows us to divide the analysis into two separate steps. In the first step, we analyze each pulsar individually and we produce posterior distributions over the Fourier coefficients describing the Gaussian processes involved in the single pulsar noise model. Then, these posteriors are combined for a GWB search over the whole array of pulsars. This method allows us to move the PTA analysis to the Fourier domain, focusing only on the signal of interest (the GWB) and those signal processes covariant with it (pulsars intrinsic red noise, DM variations, etc.). The key advantage of this method is that it can be applied to both gamma-ray and radio data, independently of the package used to interpret the timing data and build the signal model. Here, we present a brief summary of the derivation of the regularized likelihood formulation. More details can be found in Ref. [25].

For a single pulsar, marginalizing the likelihood function $p(\delta\mathbf{t}|\mathbf{b},\boldsymbol{\eta},\boldsymbol{\rho}) = \mathcal{N}(\delta\mathbf{t}|M\boldsymbol{\xi} + F\mathbf{a}, N)$ over the timing model ephemerides offsets, we obtain

$$\begin{aligned} p(\delta\mathbf{t}|\mathbf{a},\boldsymbol{\eta}) &= \int \mathrm{d}\boldsymbol{\xi}\mathcal{N}(\delta\mathbf{t}|M\boldsymbol{\xi} + F\mathbf{a}, N)p(\boldsymbol{\xi}) \\ &= \frac{\exp\left[\frac{1}{2}(\delta\mathbf{t} - F\mathbf{a})^T\tilde{N}^{-1}(\delta\mathbf{t} - F\mathbf{a})\right]}{\sqrt{\det(2\pi N)}\sqrt{\det(2\pi M^T N^{-1} M)}}, \end{aligned} \quad (4)$$

where $\tilde{N}^{-1} = N^{-1} - N^{-1}M(M^T N^{-1} M)^{-1}M^T N^{-1}$ and $p(\boldsymbol{\xi})$ is an improper prior.

For a set of hyperparameters $\boldsymbol{\rho}_0$, the product between Eq. (4) and a regression prior $p(\mathbf{a}|\boldsymbol{\rho}_0)$ can be rewritten as a normal distribution of the Fourier coefficients $\mathbf{a}$, as follows:

$$p(\delta\mathbf{t}|\mathbf{a},\boldsymbol{\eta})p(\mathbf{a}|\boldsymbol{\rho}_0) \propto \mathcal{N}(\mathbf{a}|\hat{\mathbf{a}}'_0, \Sigma'_0), \quad (5)$$

where the mean $\hat{\mathbf{a}}'_0$ and variance $\Sigma'_0$ are analytically defined for each set of values $\boldsymbol{\eta}'$ for the $\boldsymbol{\eta}$ parameters [remember that $\tilde{N} = \tilde{N}(\boldsymbol{\eta})$], as follows:

[2]Throughout the paper, we will use the notation $\mathcal{N}(\mathbf{x}|\boldsymbol{\mu}, \Sigma)$ to indicate normal distributions in $\mathbf{x}$ with mean $\boldsymbol{\mu}$ and variance $\Sigma$: $\mathcal{N}(\mathbf{x}|\boldsymbol{\mu}, \Sigma) = \exp(-0.5(\mathbf{x} - \boldsymbol{\mu})^T\Sigma^{-1}(\mathbf{x} - \boldsymbol{\mu}))/(\det(2\pi\Sigma))^{1/2}$.

$$\Sigma_0' = \left(F^T \tilde{N}^{-1} F + \phi_0^{-1}\right)^{-1}$$
$$\hat{\mathbf{a}}_0' = \Sigma_0' F^T \tilde{N}^{-1} \delta\mathbf{t}. \tag{6}$$

The matrix $\phi_0$ is the covariance matrix of the prior imposed on the Fourier coefficients distributions for $\boldsymbol{\rho} = \boldsymbol{\rho}_0$: $p(\mathbf{a}|\boldsymbol{\rho}_0) = \mathcal{N}(\mathbf{a}|0, \phi_0)$. The marginalization of the likelihood in Eq. (5) over the $\boldsymbol{\eta}$ parameters, can be approximated with great accuracy by the normal distribution

$$p(a, \boldsymbol{\rho}_0|\delta\mathbf{t}) = \int d\boldsymbol{\eta} \mathcal{N}(\mathbf{a}|\hat{\mathbf{a}}_0', \Sigma_0') p(\boldsymbol{\eta}) \approx \mathcal{N}(\mathbf{a}|\hat{\mathbf{a}}_0, \Sigma_0), \tag{7}$$

where $\hat{\mathbf{a}}_0$ and $\Sigma_0$ can be estimated numerically from samples of the $\boldsymbol{\eta}$ parameters (under the condition of fixed hyperparameters $\boldsymbol{\rho}_0$) and the corresponding $\hat{\mathbf{a}}_0'$ and $\Sigma_0'$. This concludes the first step of the GWB analysis.

In the second step, we consider the whole array of pulsars to search for correlated signals. Note that the cross-pulsar correlation information in the model is fully encoded in the distribution $p(\mathbf{a}|\boldsymbol{\rho})$. Following Eq. (7), we can write

$$p(\mathbf{a}, \boldsymbol{\rho}|\delta\mathbf{t}) p(\delta\mathbf{t}) = p(\delta\mathbf{t}|\mathbf{a}) p(\mathbf{a}|\boldsymbol{\rho}_0) \frac{p(\mathbf{a}|\boldsymbol{\rho}) p(\boldsymbol{\rho})}{p(\mathbf{a}|\boldsymbol{\rho}_0)}$$
$$\approx \mathcal{N}(\mathbf{a}|\hat{\mathbf{a}}_0, \Sigma_0) \frac{p(\mathbf{a}|\boldsymbol{\rho}) p(\boldsymbol{\rho})}{p(\mathbf{a}|\boldsymbol{\rho}_0)}, \tag{8}$$

which is simply the product of three normal distributions in $\mathbf{a}$ and the prior $p(\boldsymbol{\rho})$. The quantity $p(\delta\mathbf{t})$ is the evidence or fully marginalized likelihood. For a full array of $N_p$ pulsars, we can generalize the previous equation as

$$p(\mathbf{a}, \boldsymbol{\rho}|\delta\mathbf{t}) \approx \prod_{k=1}^{N_p} \left[\frac{\mathcal{N}(\mathbf{a}_k|\hat{\mathbf{a}}_{0,k}, \Sigma_{0,k})}{p(\mathbf{a}_k|\boldsymbol{\rho}_{0,k})}\right] p(\mathbf{a}|\boldsymbol{\rho}) p(\boldsymbol{\rho}), \tag{9}$$

where we adopt the convention that the subscript $k$ refers to the $k$th pulsar. When no subscript is present, the quantities $\mathbf{a}$ and $\boldsymbol{\rho}$ refer to the concatenation of the parameters for all the pulsars. Lastly, marginalizing over the Fourier coefficients $\mathbf{a}$, we obtain the posterior distribution for the $\boldsymbol{\rho}$ hyperparameters we are trying to find:

$$p(\boldsymbol{\rho}|\delta\mathbf{t}) \approx \frac{\mathcal{N}(\hat{\mathbf{a}}_0|0, \Sigma_0)}{\mathcal{N}(\hat{\mathbf{a}}|0, \Sigma)} \sqrt{\frac{\det(2\pi\phi_0)}{\det(2\pi\phi)}} p(\boldsymbol{\rho}), \tag{10}$$

where $\hat{\mathbf{a}}$ and $\Sigma$ are functions of $\boldsymbol{\rho}$, as follows:

$$\Sigma^{-1} = \Sigma_0^{-1} + \phi^{-1} - \phi_0^{-1}$$
$$\hat{\mathbf{a}} = \Sigma \Sigma_0^{-1} \hat{\mathbf{a}}_0, \tag{11}$$

with $\Sigma_0$ being the block-diagonal concatenation of the collection of $\Sigma_{0,k}$ of all pulsars, and $\hat{\mathbf{a}}_0$ the concatenation of all the $\hat{\mathbf{a}}_{0,k}$.

Thus, the whole inference over the GWB hyperparameters (and the hyperparameters of the signal processes covariant with it, such as intrinsic TN, DM variations, etc.) is simplified to a two-step analysis, where the final posterior distribution can be written as a normal distribution in the Fourier domain [Eq. (10)]. This method also allows us to analytically marginalize over the parameters ($\boldsymbol{\eta}$) of signals not covariant with the GWB. Such parameters include, for example, the white noise (WN) parameters, which are usually held fixed during the full array analysis of radio data. The WN modeling is another feature which is much easier to deal with in GPTA data. In fact, in GPTA data, the WN marginalization is absorbed into the marginalization over the template parameters. Most importantly, this method can be applied to both $\gamma$-ray and radio data, independently of the package used to interpret the timing data and build the signal model.

The only inputs required to evaluate the posterior in Eq. (10) are the mean $\hat{\mathbf{a}}_0$ and covariance $\Sigma_0$ for the Fourier coefficients distribution under the condition $\boldsymbol{\rho} = \boldsymbol{\rho}_0$ [Eq. (11)]. In Ref. [25], the method was tested on the radio pulsars of the EPTA DR2new dataset [20], for which $\hat{\mathbf{a}}_0$ and $\Sigma_0$ were obtained numerically from a set of samples of the $\boldsymbol{\eta}$ parameters [and the correspondent $\hat{\mathbf{a}}_0'$ and $\Sigma_0'$, as described in Eq. (6)], and the resulting posteriors for the GWB hyperparameters were equivalent to the ones obtained with the "standard" time domain formulation of the likelihood [Eq. (3)]. For gamma-ray pulsars, it is also possible to obtain samples of the Fourier coefficients describing the red signals included in the timing model.

For the analysis presented in this paper, we used SHOOGLE [31]: a Python package based on PINT [32,33] that extends the Gaussian process framework normally used in radio pulsar timing to gamma-ray pulsar data, using a Gibbs sampling approach to jointly sample timing model parameters, Fourier coefficients, noise hyperparameters, and template pulse-profile parameters.

### C. GPTA and the first data release

In 2022, the Fermi LAT Collaboration published the results of the first gamma-ray PTA experiments [24]. The dataset includes 12.5 years of observations for 35 bright gamma-ray pulsars and places a 95% credible upper limit on the amplitude of the GWB of $1 \times 10^{-14}$ at the frequency of 1 $\mathrm{yr}^{-1}$.

Two different strategies were implemented to obtain this constraint. The first method consists of folding the photon observations to obtain TOAs, and then applying the radio PTA data analysis tools [30]. However, the limited photon flux and effective collecting area require the folding of many months, up to one year, of observations to obtain a well-defined and reliable TOA. This prevents fitting for (or marginalizing over) binary parameters, since orbital periods are usually much shorter than these folding times, and potentially even for astrometric parameters. The TOA approach also requires the assumption of a single, fixed pulse profile template. A gamma-ray pulsar's intrinsic pulse shape is not known *a priori*, and so this assumption introduces
a possible systematic bias that is not accounted for in the analysis. The shape of the pulse profile determines the posterior precision on the timing parameters (a wider pulse will lead to larger timing model uncertainties), so this is equivalent to assuming fixed white noise parameters, rather than marginalizing over them. Another obstacle for the application of the TOA method to gamma-ray pulsars is the departure from Gaussianity of the uncertainties on the data points for the faint pulsars; these deviations could significantly bias the results. For the first GPTA dataset, it was only possible to compute well-defined TOAs for 29 out of the 35 pulsars.

The second approach instead does not involve any folding to estimate TOAs, but works with each photon individually: a *photon-by-photon* approach. The unbinned method uses an exact likelihood function for the set of photons arrival times, while the TOA method uses a Gaussian approximation to the exact likelihood. Despite expecting the photon-by-photon approach to have sensitivity comparable to the TOA method, the results obtained with the unbinned method are statistically more reliable and less subject to biases.

For each photon $i$, the Fermi LAT registers its energy $E_i$, the incident direction $\hat{n}_i$, and the time of arrival $t_i$ with a precision of $\sim$300 ns [17]. Because of the limited sky resolution and of the brightness of the gamma-ray sky, there is no conclusive way to assert if the observed photon was emitted by the targeted pulsar or by some background source. Thus, each photon is assigned a probability weight $0 < w_i < 1$ that the $i$th photon was emitted by the pulsar [34–37], as follows:

$$w_i = \frac{S_{\rm psr}(E_i, t_i, \hat{n}_i)}{S_{\rm psr}(E_i, t_i, \hat{n}_i) + \sum_j S_j(E_j, t_j, \hat{n}_j)}, \tag{12}$$

where $S$ are the predicted photon fluxes and $j$ runs over all the background sources. Given the weights, we can write the Poissonian likelihood for the data $D$ as a function of the timing model parameters $\boldsymbol{\beta}$, the Fourier coefficients $\mathbf{a}$ and the template pulse profile parameters $\boldsymbol{\tau}$ as

$$p(D|\boldsymbol{\beta}, \mathbf{a}, \boldsymbol{\tau}) = \prod_i w_i p(\Phi_i|\boldsymbol{\tau}) + (1 - w_i), \tag{13}$$

where $p(\Phi_j|\boldsymbol{\tau})$ is the template pulse profile (usually written as a sum of symmetrically wrapped Gaussian peaks) evaluated at the phase $\Phi_i = \Phi(t_i; \boldsymbol{\beta}, \mathbf{a})$ for the $i$th photon. The phase model $\Phi(t_i; \boldsymbol{\beta}, \mathbf{a})$ is the sum of the predicted phase from the timing model solution and the description in the Fourier basis of additional noise processes. [Note that, contrary to Eqs. (3) and (13) is not Gaussian; however, it is often a good approximation to use the quadratic expansion of the likelihood and marginalize it as if it were a Gaussian].

With the photon-by-photon approach, in the first GPTA data release single-pulsar upper limits on the GWB amplitude were inferred. Since no cross-pulsar correlated term is included in this method, the joint limit was obtained by multiplying the single pulsar posteriors and integrating the resulting distribution. In this paper, we carry out the first search for a cross-pulsar correlated signal directly from the gamma-ray photon observations.

## III. SIMULATED DATA

Before applying the method described in Sec. II B to the dataset of the first GPTA data release, we show its performance on simulated GPTA datasets. In this section, we first describe how we simulated GPTA datasets containing GWB signals of different amplitudes, and then discuss the statistical tests we carried out to infer the goodness of the inference method described in Sec. II B.

### A. Simulations setup

A gamma-ray timing dataset for a pulsar consists of a list of photon time of arrivals and their associated directions, energies, and probability weights [Eq. (12)]. To simulate GPTA datasets representative of the first GPTA dataset [24], we use the same photon weights, timing model and templates of those 35 pulsars.

Here is a brief description of our method for simulating gamma-ray pulsar photon observations: (i) For each pulsar, we randomize the order of the weights associated with the photon observations; this will be the weights list for the simulated photons. (ii) For each simulated photon observation, we sample uniformly between 0 and 1: if the sampled value is bigger than the weight associated to this photon, the photon is associated to a phase distribution uniformly sampled between 0 and 1 (*photon assigned to the background*); if the sampled value is lower than the weight associated to this photon, the phase is sampled from the normalized pulse template (*photon assigned to the pulsar*). (iii) For each photon $i$, we now have a new weight and phase $\Phi'_i$. We compute the difference in phase between the new phase and the real one observed by the telescope ($\Phi_i$), and update the time of arrival: $t'_i = t_i + (\Phi'_i - \Phi_i)/f_0$, where $f_0$ is the spin frequency of the pulsar timing solution. We thus obtain a set of noiseless[3] pulsar photon data.

To inject a GW signal into a simulated GPTA dataset, we use the same method as in Ref. [38] and implemented in the `toasim` [39] function *createGWB* for radio pulsar timing residuals. This approach produces a stationary, isotropic, and Gaussian GWB. The algorithm consists of modeling the GWB signal in Fourier space through a diagonal covariance matrix where the contributions at different frequencies are assumed to be independent, and the cross-pulsar correlations are defined by the HD curve.[4] After adding the simulated time delays to the photons of each pulsar, a refit of the spin frequency and its first derivative is necessary to readjust the initial timing solution. In fact, a TN realization has a constant, a linear and quadratic term that get added to the observed phase; thus, adjusting the spin frequency and its first derivative is enough to produce an adequate starting timing solution.

[3]By "noiseless," we mean that there is no timing noise injected in the data; the only noise present is due to the random sampling of the phase distribution.

### B. GWB inference testing

To test the accuracy of the inference of a GWB signal from a GPTA dataset through the methodology described in Sec. II B, we simulated 200 GPTA datasets following the procedure discussed in Sec. III A. Each dataset contains idealized noiseless realizations of the 35 pulsars included in the first GPTA data release published by the Fermi LAT Collaboration [24]. This means that the 35 simulated pulsars will have the same pulse template profiles, timing model parameters, and photon weight distributions as the real observed pulsars. In each dataset, we injected a GWB signal, whose log-amplitude was uniformly sampled from the interval $[-15.5, -12.5]$. No additional timing noise is included in these simulations.

We searched for the GWB amplitude in all simulated datasets using the regularized likelihood in Eq. (10). With SHOOGLE, we obtained the posteriors of the Fourier coefficients used to describe the low-frequency noise for each pulsar. The analysis was conducted over the ten lowest frequency bins, from $1/T_{\rm tot}$ (where $T_{\rm tot}$ is, in this case, 12.5 years) to $10/T_{\rm tot}$, and the power spectral density (PSD) of the Gaussian process on the individual pulsars used to define a prior on the Fourier coefficients was a flat-tail power law[5] function of frequency,

$$\phi(f; A, \gamma, \kappa) = \max\left(\frac{A^2}{12\pi^2}\left(\frac{f}{\rm yr^{-1}}\right)^{-\gamma} \rm yr^3, \kappa^2\ yr^3\right), \qquad (14)$$

where we set $\boldsymbol{\rho}_0 = [\log_{10}A_0, \gamma_0, \log_{10}\kappa_0] = [-10.2, 5, -9.2]$. We model the PSD of the GWB signal as a simple power law function of frequency, as follows:

$$\phi(f; A_{\rm gwb}, \gamma_{\rm gwb}) = \frac{A_{\rm gwb}^2}{12\pi^2}\left(\frac{f}{{\rm y}r^{-1}}\right)^{-\gamma_{\rm gwb}} \rm yr^3. \qquad (15)$$

The recovery model fixes the slope of the PSD of the GWB signal to the injected value of $\gamma_{gwb} = 13/3$, which is the theoretically predicted value for the PSD of a stochastic GWB produced by an astrophysical population of GW-driven and circular SMBHBs [40].

For each of the 200 simulated datasets, we also folded the photon observations to compute TOAs, and tested the GWB recovery from these TOAs using the standard radio methods, coded in ENTERPRISE [30] and discussed in Sec. II A. Depending on the brightness of the pulsar and the accuracy of the template fit, the cadence used for the photon folding was either 182 or 243 days.

Figure 1 shows the comparison between the posteriors for the GWB log-amplitude obtained from the 200 simulated datasets directly from the photon observations (left) and from the TOAs (right). These summary plots show that both methods yield upper limits for very low GWB amplitudes, while they allow for accurate recoveries for louder backgrounds. As expected, there is no significant difference in sensitivity between the photon-by-photon method and the TOAs method. However, the method presented here allows the template parameters to vary when collecting samples for the Fourier coefficients with SHOOGLE, and then marginalizes over them, accounting for this additional source of uncertainty. Furthermore, it allows for inclusion of the cross-pulsar correlated signals in the recovery model without having to fold the observations into TOAs (which can be problematic, especially for faint pulsars).

We also checked for possible intrinsic biases in the GWB inference from our method. Using the same 200 simulated datasets, we built a probability-probability plot (PP-plot) to test the accuracy of our recovery of the GWB amplitude (Fig. 2). A PP-plot is a statistical visualization tool to check for possible intrinsic biases in a recovery method for a specific parameter or set of parameters. The $y$-axis corresponds to the fractional number of times the recovery lies within the credible interval indicated on the $x$-axis. When the recovery is unbiased, the PP-plot curve closely follows the diagonal (see Refs. [41–43] for more details). In Fig. 2, the solid red line refers to the analyses carried out with the photon method and the regularized Fourier likelihood, while the dashed red line refers to the inference runs carried out from the TOAs. While both methods do not show signs of any alarming bias, the distribution for the photon-by-photon approach combined with the regularized likelihood formulation is more consistent with the diagonal than the distribution obtained from the TOAs analysis. The latter, in fact, is always below the diagonal of the plot and it is slightly outside the $3\sigma$ confidence interval for unbiased recoveries. Although this result does not point to any concerning bias in the recovery of the GWB amplitude from a GPTA dataset (it is comparable to the results obtained for the same study on radio PTA datasets [44–46]), it may hint at a mild tendency to underestimate the GWB amplitude.

[4]The assumption of diagonal covariance matrix between different frequencies corresponds to the ideal case of an infinite time domain support. Because of our limited observation time ($T_{\rm tot}$), this assumption does not hold and results in Gibbs phenomena in the Fourier transform. To avoid this problem, the Fourier representation is defined on a very wide and dense frequency grid [usually from $1/(10T_{\rm tot})$ up to the Nyquist frequency]. The sampled realization is then translated to the time domain through a discrete inverse Fourier transform, and linearly interpolated to the actual time of arrival of our photons.

[5]See Appendix A in Ref. [25] for a detailed discussion on the limitations on the definition of the $\boldsymbol{\rho}_0$ hyperparameter values.

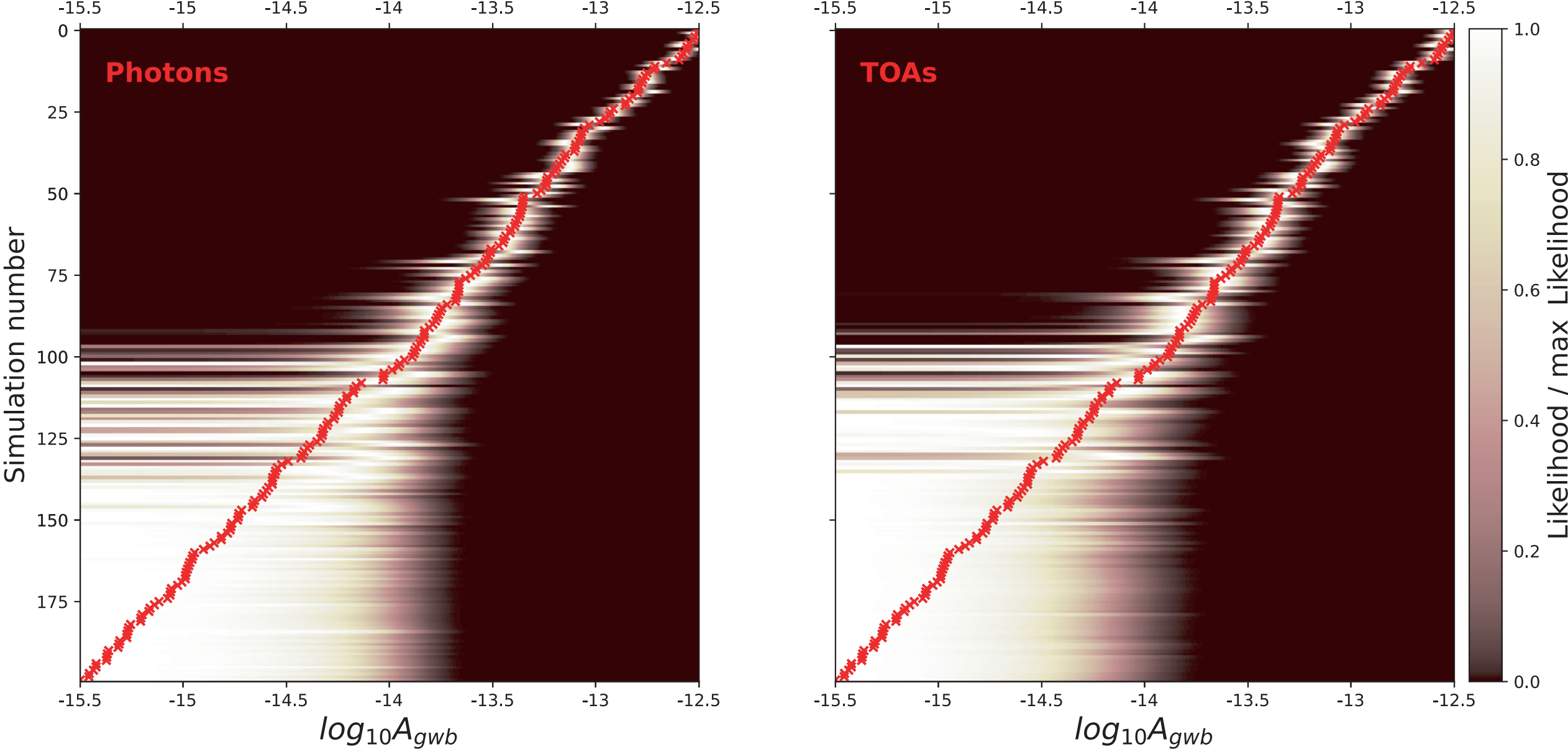

FIG. 1. Summary of the posterior distributions obtained for the 200 simulated GPTA datasets. The $x$-axis shows the prior interval of the GWB log-amplitude. Each line corresponds to a different simulation and the red crosses indicate the injected GWB amplitude. Brighter colors correspond to a higher value of the posterior distribution. It is evident from this plot that recovery gets more and more accurate for stronger simulated GWBs. Left panel: recovery directly from single photon data with the regularized likelihood method. Right panel: recovery from TOAs computed by folding the photon observations.

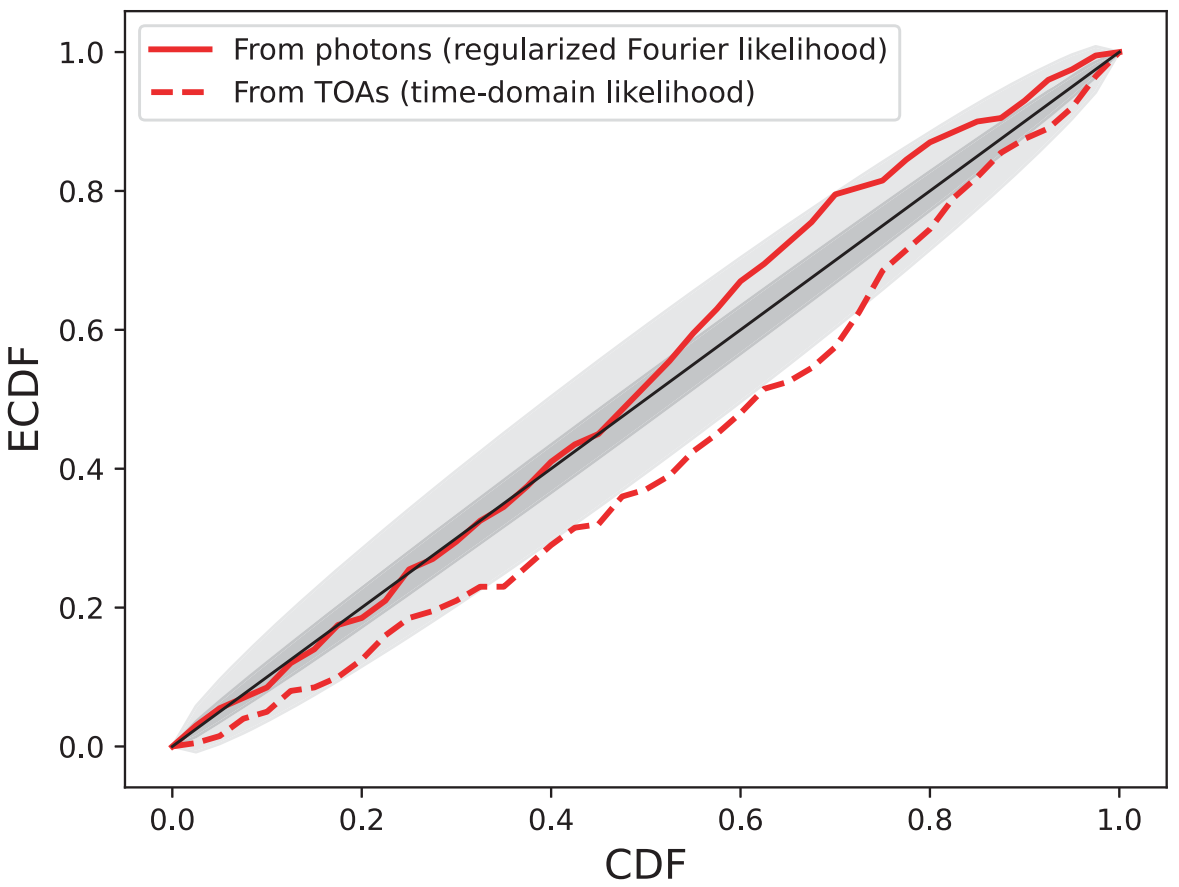

FIG. 2. Probability-probability plot for the recovery of the GWB amplitude for the set of 200 simulated GPTA datasets. The $x$-axis shows the cumulative distribution function, while the $y$-axis shows the empirical cumulative distribution function. The solid red line corresponds to the recoveries obtained directly from the photon data with the regularized likelihood method, while the dashed red line corresponds to the recovery obtained from the TOAs computed after folding the photon data. The gray areas show the $1\sigma$ and $3\sigma$ confidence intervals from the expected distribution for an unbiased recovery model (solid black thin line).

The results shown in Figs. 1 and 2 corroborate the hypothesis that the two recovery methods have comparable sensitivities to GWB signals, but the recoveries obtained with the regularized likelihood method are statistically more robust (Fig. 2).

## IV. GPTA FIRST DATA RELEASE: NEW RESULTS

We now apply the two-step analysis summarized in Sec. II B to the photon observations of the 35 pulsars in the first GPTA analysis [24]. First, for each pulsar we obtained (with SHOOGLE) the posterior distribution of the Fourier coefficients under fixed hyperparameters $\boldsymbol{\rho}_0$. Here, the recovery model included TN for each pulsar modeled over the first ten frequency bins with a flat-tail power law-shaped PSD [Eq. (14)]: $\boldsymbol{\rho}_0 = [\log_{10}A_0, \gamma_0, \log_{10}\kappa_0] = [-10.2, 5, -9.2]$. Then, using Eq. (10), we searched the full array sampling for the hyperparameters of each pulsar TN and the GWB hyperparameters. For the results presented in this paper, we model each pulsar TN over the first ten frequency bins by a (flat-tail) power law PSD, while the GWB is modeled over the lowest five frequency bins by a power law PSD.

We tested two different minimum photon weight thresholds for the photon data: (i) $w > 0.05$ for all pulsars (which is the weight threshold used in Ref. [24] for the single pulsar upper limits on the GWB amplitude); (ii) an *optimal* weight threshold for each pulsar, computed such that the sum of the squared weights for all the included photons is 99% of the sum of the squared weights for all photons:

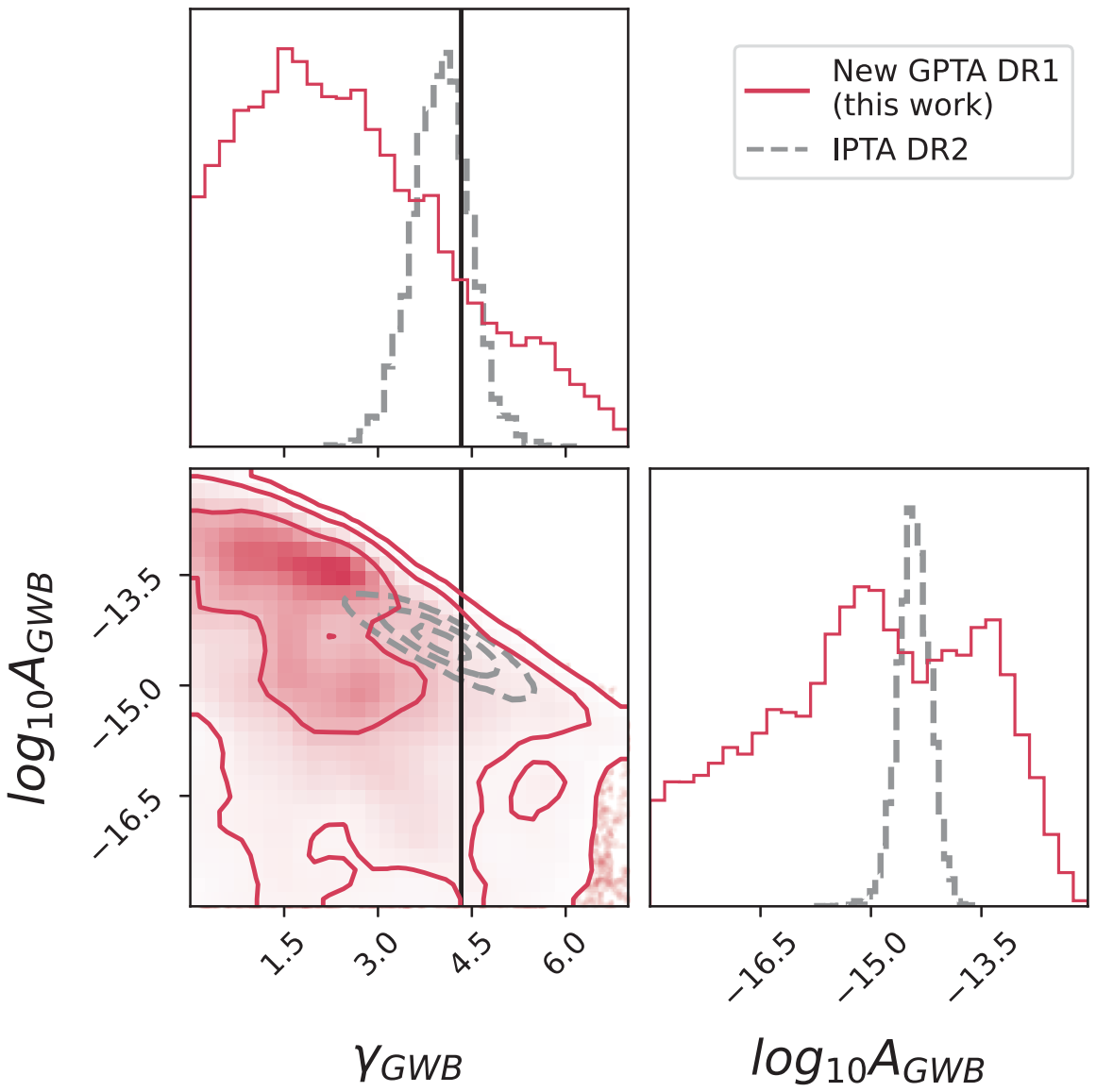

FIG. 3. GWB log-amplitude and slope posterior distribution obtained from the GPTA dataset of 35 pulsars directly from the photons with the regularized Fourier likelihood method. For each pulsar, the optimal photon weight threshold was considered (details in the text). The vertical line shows the nominal value $\gamma_{\rm gwb} = 13/3$ expected for a GWB produced by a population of SMBHBs. For comparison, the GWB posterior published in the latest IPTA data release [48] is also shown. (Note that the one-dimensional posteriors are renormalized for visualization purposes.).

$\sum_{w_i > w_{\rm opt}} w_i^2 = 0.99 \sum_i^N w_i^2$. Here, the expectation value of the H-test statistic is assumed to be roughly proportional to the sum of the weights squared; thus, this naive approximation aims at recovering nearly all of the SNR without having to accurately compute it [47]. Occasionally we find $w_{\rm opt} = 0.07$–$0.08$, but we restrict $w_{\rm opt} \leq 0.05$, which was the threshold used in the previous analysis [24].

The posteriors for the GWB hyperparameters obtained for the optimal choice of weight threshold for each pulsar are shown in Fig. 3. The posteriors are not constrained on the left side of the prior; thus, the current sensitivity allows us only to set upper limits on the amplitude of the GWB. For reference, we also show the GWB posteriors obtained from the latest International PTA (IPTA) dataset [48].

By fixing $\gamma_{\rm gwb} = 13/3$ in the analysis, the updated 95% upper limit for the GWB amplitude from the first Fermi GPTA dataset is $1.18 \times 10^{-14}$, normalized at the frequency of 1 $\rm yr^{-1}$. In Fig. 4, we show the posteriors and upper limits obtained for the GWB amplitude in the cases of the photon approach combined with regularized Fourier likelihood for two different choices of weight threshold, and for the TOAs analysis (carried out in Ref. [24]), where the TOAs are obtained folding up to one year of observation time.[6] For reference, the constrained posterior obtained from the latest IPTA dataset [48] is also shown. The updated upper limit of $1.18 \times 10^{-14}$ is very similar to the upper limit obtained from the previous analysis on the same dataset (Ref. [24], TOAs method); in fact, the two methods are not expected to have significant differences in sensitivity. However, as demonstrated with the analysis summarized in Fig. 2, this updated upper limit is, statistically, more robust.

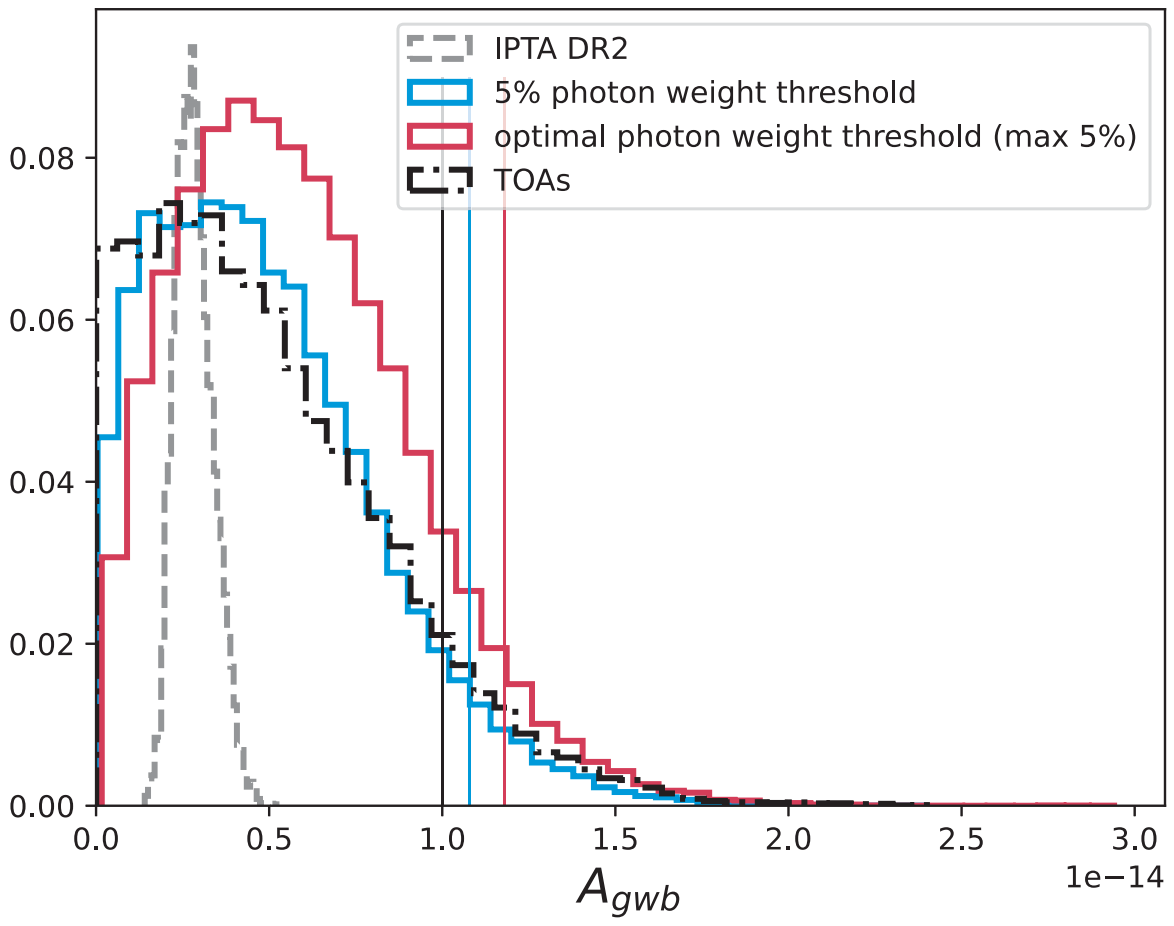

FIG. 4. GWB amplitude posteriors for $\gamma_{\rm gwb} = 13/3$ obtained for the GPTA dataset from the TOAs and from the photon-by-photon analysis using two different photon weight thresholds (the vertical lines mark the corresponding 95% confidence upper limits). The posterior published in the latest IPTA data release [48] is also shown for reference.

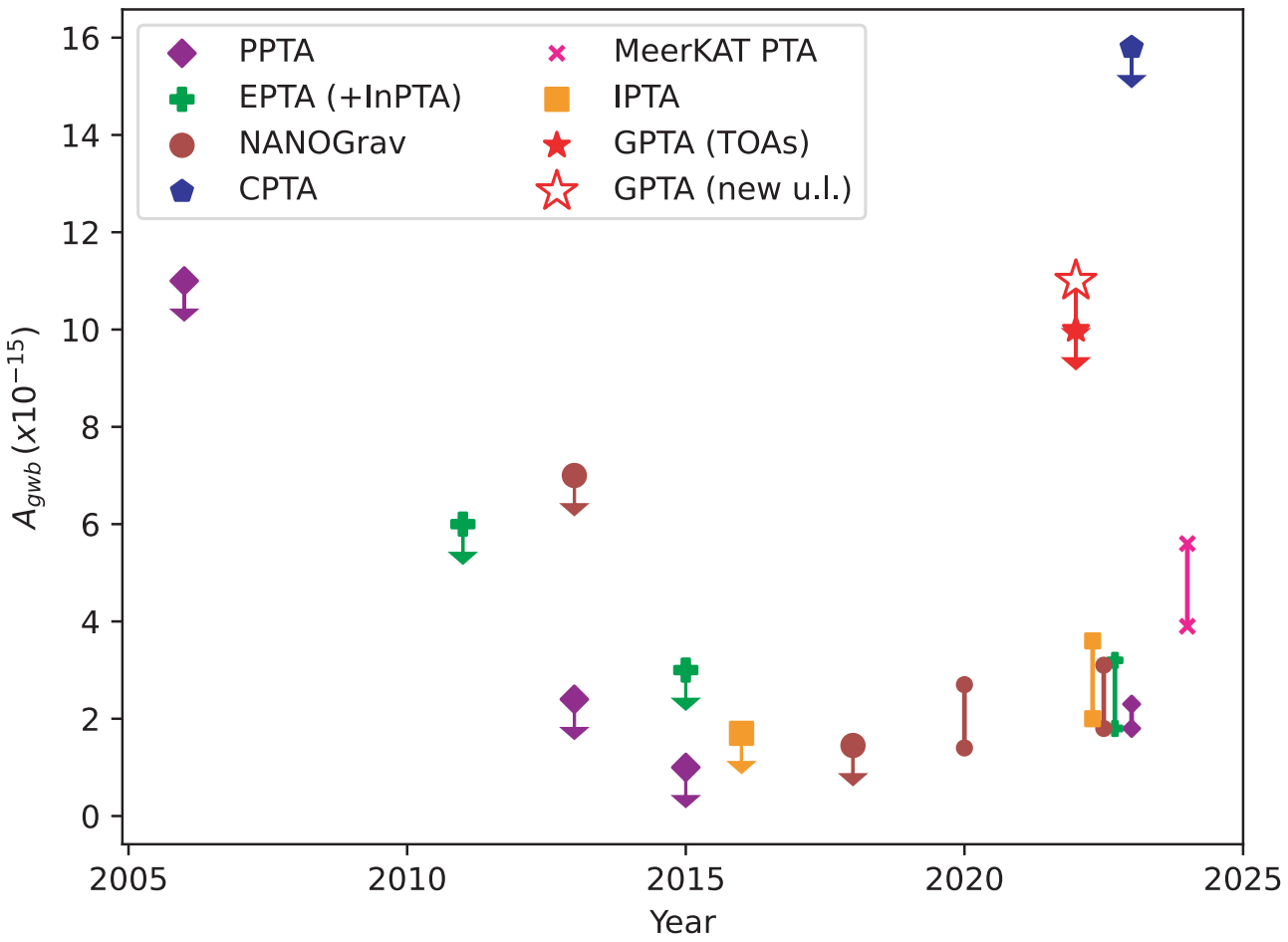

FIG. 5. Constraints on the GWB from radio and gamma-ray PTAs (updated version of Fig. 1 of Ref. [24]). The inferred constraints on the GWB amplitude at 1 $\rm yr^{-1}$ (see Table I for the references of the results shown in this figure) are plotted as a function of dataset publication date and assume $\gamma_{\rm gwb} = 13/3$, as predicted for a GWB produced by a population of supermassive black hole binaries. Upper limits at 95% confidence are shown with arrows, while amplitude ranges indicate detections of a common noise process. [Note: for clarity, the CPTA result is shown as an upper limit; see Table I for the complete confidence interval.] Recently, an additional upper limit on the GWB amplitude of $A_{\rm gwb} < 10^{-13.4}$ was obtained from x-ray pulsar observations collected by the Neutron Start Interior Composition Explorer (NICER) [49].

TABLE I. Constraints on the GWB amplitude from radio and gamma-ray PTAs, with publication references.

| Label | Reference | $A_{gwb}$ [$\times 10^{-15}$] | Range [$\times 10^{-15}$] | Note |
|---|---|---|---|---|
| PPTA 2006 | [50] | <11 | ⋯ | 95% u. l. |
| PPTA 2013 | [51] | <2.4 | ⋯ | 95% u. l. |
| PPTA 2015 | [52] | <1.0 | ⋯ | 95% u. l. |
| PPTA 2023 | [22] | 2.04 | 1.82–2.29 | 68% range |
| EPTA 2011 | [53] | <6.0 | ⋯ | 95% u. l. |
| EPTA 2015 | [54] | <3.0 | ⋯ | 95% u. l. |
| EPTA + InPTA DR2 | [20] | 2.45 | 1.74–3.16 | 95% credible region |
| NANOGrav 5-yr | [55] | <7.0 | ⋯ | 95% u. l. |
| NANOGrav 11-yr | [56] | <1.45 | ⋯ | 95% u. l. |
| NANOGrav 12.5-yr | [57] | 1.92 | 1.4–2.7 | 5%–95% credible region |
| NANOGrav 15-yr | [21] | 2.4 | 1.8–3.1 | 5%–95% credible region |
| CPTA | [19] | 2.0 | 0.025–15.8 | 5%–95% credible region |
| MeerKAT PTA | [23] | 4.8 | 3.9–5.6 | 68% range |
| IPTA DR1 | [58] | <1.7 | ⋯ | 95% u. l. |
| IPTA DR2 | [48] | ⋯ | 2.0–3.6 | 5%–95% credible region |
| GPTA (TOAs) | [24] | <10 | ⋯ | 95% u. l. |
| GPTA (photons) | [*This paper*] | <11.8 | ⋯ | 95% u. l. |

The upper limits obtained for the two different thresholds are comparable, but different: $1.09 \times 10^{-14}$ for a weight threshold of 5% for all pulsars and $1.18 \times 10^{-14}$ for the optimal weight threshold choice (setting a maximum weight cut at 5%). Between the two, the one obtained using the optimal weight threshold is expected to be more reliable, because the pulsed signal to noise ratio should be higher.

In Fig. 5 and Table I we show a summary of the constraints on the GWB amplitude from both radio and gamma-ray PTAs. We tested also different numbers of frequency bins for both pulsar intrinsic TN and the GWB; the upper limit on the GWB amplitude obtained in these tests was always comparable with the upper limit reported here.

## V. DISCUSSION AND CONCLUSIONS

We presented an updated constraint for the amplitude of the GWB signal from the 35 gamma-ray pulsars of the first GPTA data release published by the Fermi LAT Collaboration [24]. The new upper limit presented in this paper was obtained by analyzing the photons directly (rather than computing TOAs) and searching for an HD-correlated signal using the regularized formulation of the PTA likelihood in Fourier space introduced in Ref. [25].

The new upper limit for the GWB amplitude, normalized at 1 $\mathrm{yr}^{-1}$, is $1.18 \times 10^{-14}$. This is comparable to the upper limit previously published in Ref. [24] for the same dataset. This is expected. In fact, we proved with a set of 200 simulated GPTA datasets (containing GWB signals of different amplitudes) that the photon-by-photon approach combined with the regularized likelihood formulation has on average the same sensitivity as the TOAs method (Fig. 1). However, using a probability-probability plot, we showed that the TOAs method is statistically less robust for the case of the recovery of the GWB amplitude, and presents a mild bias towards underestimated values of this amplitude (Fig. 2). Furthermore, the photon-by-photon method does not fix the template parameters nor marginalizes over them, but allows for them to vary while collecting samples for the Fourier coefficient describing the TN (and GWB), leading to a more accurate (and physically correct) estimate of the noise processes involved. Nevertheless, the results shown in Fig. 2 are comparable with similar consistency checks published in the literature for GWB recoveries from radio PTAs, and do not point to any alarming bias in the analysis with both methods.

In Fig. 5 and Table I, we summarize the constraints on the GWB amplitude from both radio and gamma-ray PTAs.

Having a GPTA experiment (and, more in general, other PTA experiments using pulsar observations at higher energies) alongside the radio PTAs is of great importance and value. First of all, it provides a completely independent confirmation of the results obtained by the radio PTAs. The GPTA data have a (essentially) constant experimental setup and, thus, they do not require a large number of instrument dependent noise parameters or phase jumps. Such stability is a good advantage to model GWs at frequencies below 0.1 $yr^{-1}$, where most of the GWB astrophysical information is encoded. However, the biggest advantage of GPTA experiments is the independence (due to the high photon energy of the gamma-ray emissions) from effects such as interaction with the interstellar medium, scattering noise, solar wind, etc. Thus, dispersion measure and, consequently, DM variations are not part of gamma-ray pulsar noise models. Correctly disentangling TN and DM variations in radio PTA data is one of the biggest challenges for the current analysis. Being able to break the degeneracy between TN and DM variations would not only provide valuable information on the interstellar medium distribution, but would also significantly improve our sensitivity to a GWB signal.

[6]The TOAs posterior shown in Fig. 4 is not the original result of the analysis carried out by the authors of Ref. [24], but has been computed by the authors of this paper following the prescriptions in Ref. [24]. The results obtained are consistent with those reported in Ref. [24].

Calculations were made using NumPy [59], Matplotlib [60], CORNER plot [61], SciPy [62], ENTERPRISE [30], and PTMCMCSampler [63].

## ACKNOWLEDGMENTS

This work was supported by the Max Planck Gesellschaft (MPG) and the ATLAS cluster computing team at AEI Hannover. The Fermi LAT Collaboration acknowledges generous ongoing support from a number of agencies and institutes that have supported both the development and the operation of the LAT as well as scientific data analysis. These include the National Aeronautics and Space Administration and the Department of Energy in the United States, the Commissariat à l'Energie Atomique and the Centre National de la Recherche Scientifique/Institut National de Physique Nucléaire et de Physique des Particules in France, the Agenzia Spaziale Italiana and the Istituto Nazionale di Fisica Nucleare in Italy, the Ministry of Education, Culture, Sports, Science and Technology (MEXT), High Energy Accelerator Research Organization (KEK) and Japan Aerospace Exploration Agency (JAXA) in Japan, and the K. A. Wallenberg Foundation, the Swedish Research Council and the Swedish National Space Board in Sweden. Additional support for science analysis during the operations phase is gratefully acknowledged from the Istituto Nazionale di Astrofisica in Italy and the Centre National d'Études Spatiales in France. This work performed in part under DOE Contract No. DE-AC02-76SF00515. Work at NRL is supported by NASA under the Fermi Guest Investigator Program, No. NNG22OB35A. S. V., A. C., and A. P. acknowledge financial support from the European Research Council (ERC) starting grant GIGA (Grant Agreement No. 101116134). A. P. acknowledges financial support through the NWO-I Veni fellowship. We acknowledge financial support from the "Programme National de Cosmologie et Galaxies" (PNCG), "Programme National Hautes Energies" (PNHE) and "Programme National Gravitation, Références, Astronomie, Métrologie" (PNGRAM) of CNRS/INSU, France. We acknowledge financial support from Agence Nationale de la Recherche (ANR-18-CE31-0015), France. We would like to thank David Smith for reviewing this paper on behalf of the Fermi LAT Collaboration.

## DATA AVAILABILITY

The data that support the findings of this article are openly available [64].